\documentclass{AIAA}
\usepackage{amsmath,amsthm,amssymb,graphicx,bm,url}
\usepackage{color}
\usepackage{algorithmic}
\usepackage{graphicx}
\usepackage{textcomp}
\usepackage{graphicx,bm,url}
\usepackage{algorithm2e}
\usepackage{blkarray, bigstrut}

\newcommand{\MtoT}{\text{M-to-T}}
\newcommand{\TtoM}{\text{T-to-M}}

\newcommand{\MtoE}{\text{M-to-E}}
\newcommand{\EtoM}{\text{E-to-M}}
\newcommand{\std}{\text{std}}
\newcommand{\Ref}{\text{ref}}
\newcommand{\RTN}{\text{RTN}}

\newcommand{\CRTN}{}
\newcommand{\tCRTN}{}
\newcommand{\obs}{\text{obs}}
\newcommand{\blank}[1]{\hspace*{#1}}
\newcommand{\nil}[1]{}
\newcommand{\atantwo}{\text{atan2}}

\begin{document}

\title{Revisiting the orbital tracking problem}

\author{John T. Kent\footnote{Professor, Department of Statistics, j.t.kent@leeds.ac.uk.} and Shambo Bhattacharjee\footnote{PhD student, Department of Statistics, mmsb@leeds.ac.uk.}}
\affiliation{University of Leeds, Leeds, UK}
\author{Weston R. Faber\footnote{Research Scientist, weston.faber@l3harris.com.}}
\affiliation{L3Harris, Applied Defense Solutions, Colorado Springs, CO, USA}
\author{Islam I. Hussein\footnote{Senior R\&D Scientist, islam.Hussein@l3harris.com.}}
\affiliation{L3Harris, Applied Defense Solutions, Herndon, VA, USA}

\begin{abstract}
Consider a space object in an orbit about the earth. An uncertain
initial state can be represented as a point cloud which can be
propagated to later times by the laws of Newtonian motion.  If the state
of the object is represented in Cartesian earth centered inertial
(Cartesian-ECI) coordinates, then even if initial uncertainty is
Gaussian in this coordinate system, the distribution quickly becomes
non-Gaussian as the propagation time increases.  Similar problems arise
in other standard fixed coordinate systems in astrodynamics,
e.g. Keplerian and to some extent equinoctial.  To address these
problems, a local ``Adapted STructural (AST)'' coordinate system has
been developed in which uncertainty is represented in terms of
deviations from a ``central state''.

Given a sequence of angles-only measurements, the iterated nonlinear
extended (IEKF) and unscented (IUKF) Kalman filters are often the most
appropriate variants to use. In particular, they can be much more
accurate than the more commonly used non-iterated versions, the extended
(EKF) and unscented (UKF) Kalman filters, especially under high
eccentricity.  In addition, iterated Kalman filters can often be
well-approximated by two new closed form filters, the
observation-centered extended (OCEKF) and unscented (OCUKF) Kalman
filters.
\end{abstract}

\maketitle

\section{Introduction}
\label{sec:intro}

This paper revisits the filtering problem for an object in orbit
about the earth.  The aim is to give a \emph{better solution} than
earlier authors.  Recall the orbital filtering problem is nonlinear:
depending on the coordinate system, either the system equation or the
observation equation (or both) is nonlinear, thus complicating the
implementation of the propagation and/or update equations in the
filter.  The  proposed solution uses a coordinate system in which the
filtering problem almost reduces to a linear Gaussian system, so that
a mild variant of the classic Kalman filter can be used.  In
particular, the update equations have a simple (nearly) closed form
and are (nearly) optimal.

For simplicity in this paper, assume the following 
\emph{idealized conditions}:

\begin{itemize}
\item[(a)] initial Gaussian uncertainty in Cartesian-ECI coordinates,

\item[(b)] Keplerian dynamics for the evolution of the state of the
  object, and

\item[(c)] a sequence of angles-only measurements made by an ideal
  observer at the center of the earth following a concentrated Fisher
  distribution on the unit sphere.
\end{itemize}

Attention is limited to idealized conditions in this paper so that  the
key features of our solution can be clearly understood.  Assumption
(a) is a common assumption in this area.  The extension of assumptions
(b) and (c) to more realistic conditions (e.g. to accommodate the
perturbation effects of the earth's gravity and to locate observers on
the surface of the earth) will be
dealt with elsewhere.

The essence of the proposed procedure can be summarized as follows:

\begin{itemize}
\item[(a)] Represent the uncertainty in a new set of coordinates which
  are called  AST coordinates.  These coordinates are essentially a local
  version of equinoctial coordinates with some small modifications.
  In these coordinates the system equation is \emph{exactly linear}.

\item[(b)] Given an observation, carry out the update step of the
  filter using the iterated unscented Kalman filter (IUKF)~\cite{b22}.
\end{itemize}

Orbital uncertainty propagation and orbital object tracking are a key
themes in Space Situational Awareness (SSA) and a number of papers
have been published in recent years to deal with the nonlinearity of
the system equation when expressed in Cartesian-ECI coordinates.
There are two basic strategies to deal with  nonlinearity: (i)
transform the coordinate system to remove the nonlinearity, or (ii) develop
sophisticated methods to accommodate it.  As an example of the former
approach, Junkins, Akella, and Alfriend~\cite{b1} showed using point
clouds that Gaussianity is approximately preserved under propagation
in equinoctial coordinates in many circumstances. The current paper
can be viewed as an extension of that point of view.

On the other hand, many other papers have taken the second approach.
For example, Park and Scheeres~\cite{b5} used a mixture (hybrid
approach) of a simplified dynamic system (SDS) model and the state
transition tensor (STT) model to propagate and model the uncertainty
with higher order Taylor series terms~\cite{b5, b12, b13}.  Vittaldev,
Russell and Linares~\cite{b6} proposed a mixture of polynomial chaos
expansion (PCE) and Gaussian Mixture Models (GMMs) based on Hermite
polynomials.  Several other papers~\cite{b15, b16} also used the
polynomial chaos model (PCM) and PCE for representing orbital
uncertainty. Horwood and Poore~\cite{b2} proposed a Gauss Von Mises
(GVM) filter using second order trigonometric terms.  

The paper is organized as follows.  Section \ref{sec:kepler} briefly
reviews some key ideas in orbital dynamics under Keplerian dynamics.
Section \ref{sec:coordinates} recalls some standard coordinate systems for
the state vector and uses them to motivate the new AST coordinate system.  The
approximate linearity between Cartesian and AST coordinates at an
initial time is explored in Sections \ref{sec:first}
and \ref{sec:linearity}.  The filtering problem using AST coordinates
is analyzed in Section \ref{sec:filtering} for a sequence of angles
only measurements.  It is demonstrated that a suitably formulated IUKF
is nearly optimal under a wide range of conditions.

\section{Review of Keplerian dynamics}
\label{sec:kepler}
\subsection{Three angles in orbital dynamics}
\label{sec:kepler:three}
A small object orbiting the earth follows an exact elliptical orbit
under Keplerian dynamics, with the center of the earth at one of the
focal points of the ellipse.  There are three angles of mathematical
interest in this setting to describe the angular position of the
object along its orbit: the \emph{eccentric anomaly} ($E$), the
\emph{mean anomaly} ($M$) and the \emph{true anomaly} ($T$), where all
three angles are measured from perigee. The true anomaly describes the
actual angular position of the object, as measured from the center of
the earth.  The mean anomaly simplifies the mathematical development
because it changes at a constant rate in time, and the eccentric
anomaly is an intermediate angle of no direct interest.  The relation
between the angles is given as follows~\cite{b4}, where $e$ is the
ellipticity, $0 \leq e < 1$:
\begin{align*}
\tan \frac12 T &= \sqrt{\frac{1+e}{1-e}} \tan \frac12 E,\\
M &= E - e \sin E.
\end{align*}
These mappings are bijective, so any one angle determines the other
two.  The calculations are all straightforward, except that a
numerical iteration is needed to solve for $E$ from $M$.

Initially all three angles are defined on the same interval
$-\pi \leq E,M,T \leq \pi$.  The angles agree at the midpoint and
endpoints.  That is, if $E=0,\pi$ or $-\pi$, then $M$ and $T$ also
equal to $0,\pi$ or $-\pi$, respectively.  Further the identification
between angles is symmetric about the origin.  That is, if $E$
corresponds to $M$ and $T$, then $-E$ corresponds to $-M$ and $-T$.
Finally, by periodic extension, the mapping between the three angles
can be extended to any interval
$-\pi + 2\pi k \leq E,M,T \leq \pi +2 \pi k$, $k \in \mathbb{Z}$.

The notation $E = F_\MtoE(M,e)$ is used to describe the
transformation between $M$ and $E$ and similar notation for the
transformations between other pairs of angles.  The main
transformations of interest are $F_\MtoE$ and
$F_\EtoM$.

\subsection{Equations of orbital motion}
\label{sec:kepler:equations}
Consider the \emph{state} of an object orbiting the earth.  The state
at time $t$ can be described in Cartesian-ECI coordinates by three-dimensional position and velocity
vectors $\bm x(t),\ \dot{\bm x}(t)$, say.  The state at any one time
determines the state at all other times under Keplerian dynamics, and the object follows an
elliptical orbit. Various features can be extracted from the state to
help describe this elliptical orbit~\cite{b4, b25, b26, b27}.  Here $\mu$ is the gravitational
constant for the earth.

(a) In general, the term \emph{basis} will be used in the paper as shorthand for a
positively oriented basis of three orthonormal vectors in
$\mathbb{R}^3$.  In the current setting, a useful basis is the
\emph{RTN (radial-tangential-normal)} basis at an initial time $t=0$,
defined as follows:
\begin{align}
\bm u &= \bm u^\RTN \propto  \bm x(0), \notag \\ 
\bm v &= \bm v^\RTN \propto \dot{\bm x}(0)-\{\dot{\bm x}(0)^T \bm u\}\bm u, 
\label{eq:jtk-rtn-defn}
\\
\bm w &= \bm w^\RTN= \bm u \times \bm v \propto \bm x(0) \times \dot{\bm x}(0), \notag
\end{align}
so that $\bm u$ points in the radial direction, $\bm v$ points in the
tangential direction (after orthogonalizing with respect the the
radial direction) and $\bm w$ is normal to the $\bm u - \bm v$ plane.
This basis is positively oriented since
$\text{det}[\bm u \ \bm v \ \bm w] = +1$, not $-1$.

(b) The \emph{angular momentum vector} is given by
$\bm h = \bm x(0) \times \dot{\bm x}(0) = h {\bm w}$, where magnitude $h = |\bm h|$
is called the \emph{angular momentum}.  

(c) The \emph{ellipticity vector} is given by
$$
\bm e = \frac{1}{\mu} (\dot{\bm x}(0) \times \bm h) -\bm u.
$$
Its  magnitude $e = |\bm e|$ is called the \emph{ellipticity}, and for this paper it is assumed that  $0 \leq e <$ 1. 

(d) The length of the  \emph{semi-major axis} of the ellipse (i.e., the
arithmetic average of the radius of the orbit at perigee and apogee)
is given by
\begin{equation}
\label{eq:a}
a = \frac{h^2/\mu}{1-e^2}.
\end{equation}

(e) The \emph{period} and the \emph{mean motion} are
\begin{equation}
\label{eq:p}
p = 2\pi \sqrt{a^3/\mu}, \quad
n = 2\pi/p = \sqrt{\mu/a^3}.
\end{equation}

(f) The \emph{direction of perigee} is given by 
$$
\theta_p = \text{atan2}(\bm e^T \bm v, \bm e^T \bm u)
$$
and defines the angle in the $\bm u - \bm v$ plane at which the
orbiting object is closest to the earth.  Here atan2 is the
two-argument arctan function found in many computing languages.  For
example, $\theta_p=0$ points towards the positive $\bm u$ axis and
$\theta_p = \pi/2$ points towards the positive $\bm v$ axis.

The ellipticity vector lies in the $\bm u - \bm v$ plane and can
be written in the form
\begin{equation}
\label{eq:f}
\bm e = f_1 \bm u + f_2  \bm v,
\end{equation}
where $f_1 = e \cos \theta_p$ and $f_2 = e \sin \theta_p$.

The Cartesian-ECI coordinates of the state can be computed from any 6
functionally independent features, such as the following:

\begin{equation}
\begin{split} 
\label{eq:features}
&\text{the $\bm u,\bm v, \bm w$ basis, (3 degrees of freedom)}\\
&\text{the ellipticity coordinates $f_1,f_2$ (2 degrees of freedom)}\\
&\text{the mean motion $n$ (one degree of freedom).}
\end{split} 
\end{equation}
Note this information does not yet form a coordinate system because we
have not yet given an explicit representation for the vectors
$\bm u,\bm v, \bm w$.


The preceding information can be used to describe the evolution of an
orbiting object in time  under Keplerian dynamics.  Note the basis 
$\bm u,\bm v, \bm w$ is defined at  time $t=0$ and so does not change with time.
In addition, the  features $f_1,f_2, n$, and hence also
$h, e, n, \theta_p$ are also constant in time.
The state equation of the orbiting object can be expressed as 
\begin{equation}
\label{jtk-eq:rtn-radial}
\bm x(t) = r(t) \{\cos \theta(t) \bm u + \sin \theta(t) \bm v\},
\end{equation}
in terms of a radial function $r(t)$ and an angular function $\theta(t)$,
where 
\begin{align}
r(t) &= (h^2/\mu)/
\{1 + f_1 \cos \theta(t) + f_2 \sin \theta(t)\} \notag\\
&= (h^2/\mu)/\{1 + e \cos (\theta(t)-\theta_p)\} \label{eq:radial}\\
\theta(t) &= \theta_p +F_\MtoT(\phi(t)-\phi_p,e),
\label{eq:angular}\\
\phi(t) &= \phi_p + F_\TtoM(\theta(0)-\theta_p,e)+n t \notag\\
&= \phi(0) + n t.
\label{eq:mean-to-true}
\end{align}
Here $\phi(t)$ denotes the propagated angle on the mean anomaly scale,
and $\theta(t)$ denotes the propagated angle on the true anomaly
scale, initialized so that $\phi(0)= \theta(0) = 0$.  Similarly,
$\phi_p = F_\TtoM(\theta_p,e)$ denotes the direction of perigee on the
mean anomaly scale where $\theta_p$ denotes the corresponding value on
the true anomaly scale.

Equation (\ref{eq:mean-to-true}) shows that on the mean anomaly scale
the angular speed $n$ is constant.  However, nonlinear mappings, centered at
the direction of perigee, are needed to move back and forth between
the mean anomaly and true anomaly scales (equation
(\ref{eq:angular})). The Roman letters $T(t)$ and $M(t)$ are used to denote
the true and mean anomalies; these angles vanish at perigee.  On the other
hand, the Greek letters $\theta(t)$ = $T(t) - T(0)$ and $\phi(t)$ = $M(t) - M(0) = nt$ are used to are used to denote
the true angular position and its conversion to the mean anomaly scale,
initialized so that $\theta(0) = \phi(0) = 0$.

The RTN basis can be contrasted to two related bases:

(a) the \emph{perifocal basis} for which the basis vectors in the
$\bm u-\bm v$-plane are given by the direction of perigee and its rotation
by $90^o$.  The perifocal convention is unsuitable for
a circular orbit (or near-circular orbit) because in this
case the direction of perigee is not defined (or is poorly defined),
leading to irrelevant ``noise'' in later calculations.

(b) the \emph{RIC (radial, in-track, cross-track) basis} for which the
basis vectors in the $\bm u-\bm v$-plane vary with the time; for each
$t$ they point in the radial and tangential directions.  Thus the RTN
basis is essentially a static version of the RIC basis, after setting
the time argument to be the initial time $t=0$.  For our purposes it is
more helpful to represent the angular position of the object along its orbit
separately from the underlying basis.

\section{Coordinate systems}
\label{sec:coordinates}
\subsection{Cartesian-ECI coordinates}
\label{sec:coordinates:eci}
In \emph{Cartesian earth centered inertial (Cartesian-ECI)} coordinates, the
position of an object orbiting the earth is represented with respect
to the \emph{standard ECI basis}, 
\begin{equation}
\label{eq:std-basis}
\bm u^\std = \begin{bmatrix} 1 \\ 0 \\ 0  \end{bmatrix}, \quad
\bm v^\std = \begin{bmatrix} 0 \\ 1  \\ 0  \end{bmatrix}, \quad
\bm w^\std = \begin{bmatrix} 0 \\ 0 \\ 1 \end{bmatrix},
\end{equation}
where $\bm w^\std$ points towards
celestial north, $\bm u^\std$ points towards the vernal direction,
and $\bm v^\std$ is rotated $90^o$ east about the celestial equator.

Cartesian-ECI coordinates are the simplest coordinates in which to
represent the state (i.e. position and velocity) of an orbiting object,
$(\bm x(t), \ \dot{\bm x}(t))$, say.  In may ways Cartesian-ECI
are the easiest coordinates to work with, but they have two main
drawbacks.  First, the propagation equations are nonlinear, leading to
non-Gaussian distributions under propagation (see the next section).
Secondly, the orbital invariants are not obvious in these coordinates.

Hintz ~~\cite{b7} provides a concise summary of different coordinate
systems for orbital dynamics.  Two of these coordinate systems 
are described next.

\subsection{Keplerian elements}
\label{sec:coordinates:kepler}
\emph{Keplerian elements} form a coordinate system 
which picks out the orbital invariants explicitly.  To define these
elements, it is necessary to start with a \emph{reference basis} of
$\mathbb{R}^3$, specified by three orthonormal vectors
$\bm u^\Ref, \bm v^\Ref, \bm w^\Ref$, say, with a positive
orientation.  Conventionally, the \emph{standard reference basis} from 
(\ref{eq:std-basis})
is used for the definition of Keplerian elements, but to motivate
later sections, a more general choice is allowed in this section.

The first two basis vectors $\bm u^\Ref$ and $\bm v^\Ref$ define a
\emph{reference plane} (i.e. the ``equatorial'' plane).  In this plane
a \emph{preferred reference direction} is given by $\bm u^\Ref$ and
angles in the reference plane can be measured counter-clockwise
(i.e. moving from $\bm u^\Ref$ to $\bm v^\Ref$) from the preferred
reference direction.  Similarly, $\bm w^\Ref$ can be termed the
\emph{normal reference direction}.

The state of the orbiting object determines a second plane, the
\emph{orbital plane}.  Within the orbital plane, a \emph{preferred
  orbital direction} can be  defined by the direction of perigee, and
angles in the orbital plane can be measured counter-clockwise (i.e. in
the direction of orbital motion) from the preferred orbital direction.

The intersection between the reference plane and the orbital plane
determines a line called the \emph{node line}.  This line determines
two opposite directions; the RAAN direction is given by the direction
for which the projection of the velocity of the orbiting object onto
the normal reference direction is positive.  Here RAAN stands for
Right Ascension of Ascending Node.  It is helpful to distinguish
between the RAAN \emph{direction}, a unit vector lying on the node
line, and the RAAN \emph{angle} $\Omega$, or RAAN for
short, specified below.

The Keplerian elements for a state ($\bm x(t), \ \dot{\bm x}(t)$),
defined with respect to the specified reference basis, are given as follows:
\begin{itemize}
\item $i=\cos^{-1} (\bm w^T \bm w^\Ref) \in [0, \pi]$, the
  inclination of the orbital plane with respect to the reference
  plane.  Here $\bm w$ denotes the unit normal vector to the orbital
  plane.
\item $\Omega \in [0, 2\pi)$, the Right Ascension of Ascending Node
  (RAAN), i.e. the angle in the reference plane from the preferred
  reference direction to the RAAN direction.
\item $e$, the eccentricity, $0\leq e < 1$
\item $\omega \in [0,  2\pi)$, the argument of perigee, i.e. the
angle in the orbital plane from the RAAN direction to the
preferred orbital direction.
\item $a$, the major semi-axis, $a >0$
\item $T(t)$, the true anomaly.
\end{itemize} 
Figure \ref{fig1001} illustrates the key directions and angles.  Note
that the Keplerian elements, except the last one, do not vary with
time.  Three of these elements, $i,\Omega, \omega$, depend on
the choice of reference plane; the other three elements, $e,a,T(t)$, do not. 

Unfortunately, Keplerian elements are not generally suitable for
representing uncertainty by Gaussian distributions.  In particular,
(a) the RAAN angle $\Omega$ becomes undetermined when the orbital
plane is either prograde equatorial ($i=0$) or retrograde
equatorial ($i=\pi$), and (b) the argument of perigee
($\omega$) becomes undetermined for a circular orbit ($e=0$).

\subsection{Equinoctial elements}
\label{sec:coordinates:equi}
The problems of Keplerian elements are partly resolved by using a
third coordinate system, \emph{equinoctial elements}, denoted
$E_1, \ldots, E_6$ and defined as follows (with respect to the same
standard reference basis that used above  for Keplerian
elements):

\begin{alignat}{3}
E_1 &= 2\tan (i/2) \cos (\Omega), & \quad
E_2 &= 2\tan (i/2) \sin (\Omega), & \quad
E_3(t) &= \Omega + \omega + T(t), \notag\\
E_4 &= e \cos (\Omega + \omega), & \quad
E_5 &= e \sin (\Omega + \omega), & \quad
E_6 &= a. \label{eq:equinoctial}
\end{alignat}

Some authors replace the major semi-axis $a$ with another quantity
depending on the ``size'' of the ellipse, e.g. the mean motion
$n$~\cite{b9, b18}.  As indicated in the notation, all the coordinates
are fixed in time except the third, and $E_1--E_5$ depend on the reference
basis.

The third coordinate $E_3(t)$ can be termed the \emph{remapped angular
  position} or \emph{break angle}~\cite{b28}.  If the orbital plane is
rotated about the node line onto the reference plane, then $E_3(t)$
represents the angular distance in the reference plane between the
preferred reference direction and the image of the position of the
orbiting object.

\begin{figure}[t]
\begin{center}
\includegraphics[width=8.2cm, keepaspectratio]{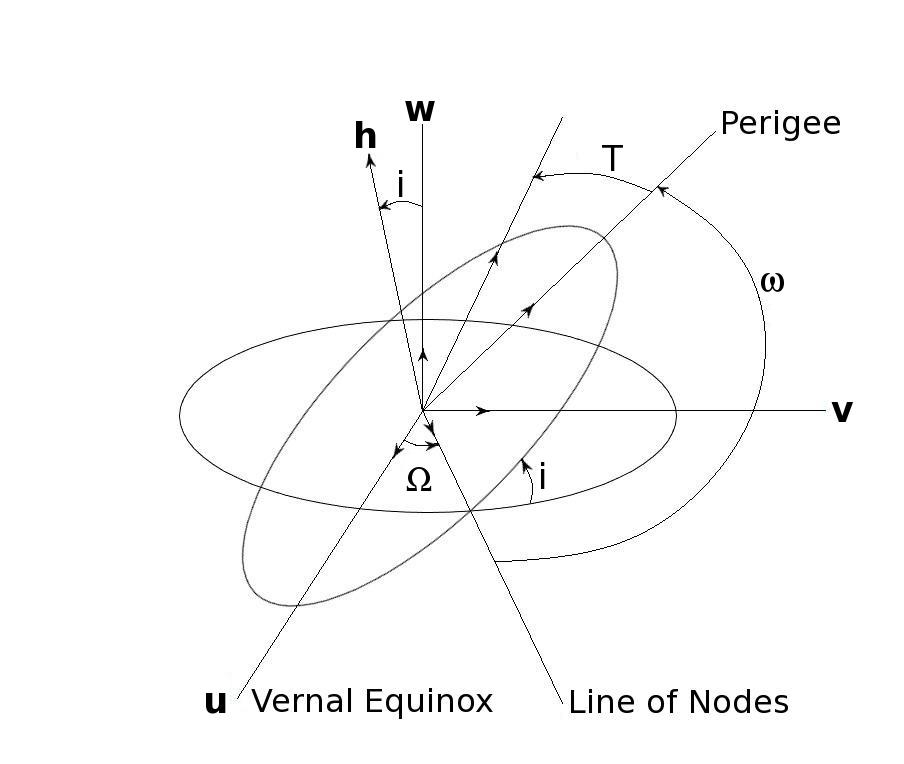} 
\caption{\textbf{Orbital dynamics.} An orbiting object in Keplerian
  dynamics. The reference directions
  $\bm u=\bm u^{\text{ref}},\bm v=\bm v^{\text{ref}},\bm w=\bm
  w^{\text{ref}}$ and the angles $\omega$, $\Omega$, $i$ and
  $T = T(t)$ are highlighted.}
\label{fig1001} 
\end{center}
\end{figure}

Even though $\Omega$ and/or $\omega$ may be undetermined in certain
circumstances, the equinoctial elements remain well-defined (except
for a retrograde equatorial orbit).  For example, for a prograde
equatorial orbit, $E_1$ and $E_2$ equal 0 since $i=0$, regardless of
the fact that $\Omega$ is undetermined.  Similarly, for a circular
orbit $E_4$ and $E_5$ equal 0 since $e=0$.  For the remapped angular
position $E_3$, it is helpful to refer to Figure \ref{fig1001}.  It
can be checked that, except for a retrograde orbit, the sum of angles
$\Omega + \omega + T(t)$ remains well-determined even though some of
the individual terms become undetermined.

The one situation where equinoctial coordinates break down is for a
retrograde orbit ($i= \pi)$. The problem can be partly resolved
by using one of two possible reference planes, namely the equatorial
plane with an upwards or downwards normal, respectively, depending on
the orbit to be described (e.g. Cefola~\cite{b20}), but the solution
below using AST coordinates can be applied more automatically and
offers more insights.

\subsection{AST coordinates}
\label{sec:coordinates:ast}

The final coordinate system to be introduced in this section is the
new \emph{Adapted STructral (AST) coordinate system}.  The philosophy
is somewhat different to the fixed coordinate systems described above.
Instead, a \emph{local} or \emph{adapted} coordinate system is used, 
designed so that uncertainty in the state can generally be
represented using Gaussian distributions.  

The starting point for AST coordinates is a known approximate value
for the state of the orbiting object at time $t=0$.  This value is
called the \emph{central state}
$(\bm x^{(c)}(0),\dot{\bm x}^{(c)}(0))$.  The purpose of the central
state is to provide a reference basis.  Its exact choice does not
matter.  In general, features related to the central state will be
indicated with a superscript $^{(c)}$.

Uncertainty in the initial state vector is represented by a notional
point cloud of \emph{deviated states} lying near the central state,
with a typical deviated state denoted $(\bm x(0),\dot{\bm x}(0))$.  In
practice the distribution of the deviated states will be modelled by a
6-dimensional Gaussian distribution.  The central state will often lie
near the middle of the point cloud, but the exact choice does not
matter.

Let CRTN stand for the \emph{central RTN basis}, that is, the RTN
basis determined from the central state at the initial time $t=0$.
Then AST coordinates, denoted $A_1, \ldots, A_6$, are defined to be a
local version of equinoctial coordinates, that is, the equinoctial
coordinates defined with respect to the CRTN basis, with some small
adjustments:

\begin{alignat}{3}
A_1 &= 2\tan (i\tCRTN/2) \cos (\Omega\tCRTN), & \quad
A_2 &= 2\tan (i\tCRTN/2) \sin (\Omega\tCRTN), & \quad
A_3(t) &= \phi\tCRTN(t), \notag\\
A_4 &= e \cos (\theta_p\tCRTN), & \quad
A_5 &= e \sin (\theta_p\tCRTN), & \quad
A_6 &= n, \label{eq:ast}
\end{alignat}
where the various angles are defined by
\begin{align}
\theta\tCRTN(t) &= \Omega\tCRTN+\omega\tCRTN+T(t) \label{eq:theta-crtn}\\
\theta_p\tCRTN &= \Omega\tCRTN+\omega\tCRTN \label{eq:thetap-crtn}\\
\phi_p\tCRTN &= F_\TtoM(\theta_p\tCRTN,e) \label{eq:phip-crtn}\\
\phi\tCRTN(t) &= \phi_p\tCRTN + M(t) \label{eq:phi-crtn}
\end{align}
Note that $A_3(t)$ is the only AST coordinate that varies with time;
the others are fixed under Keplerian dynamics.

Conversely, the Keplerian elements with respect the CRTN basis can
be defined in terms of AST coordinates by
\begin{align}
e &=\sqrt{A_4^2+A_5^2}, \quad \theta_p = \atantwo(A_5,A_4), \notag\\
i &= 2 \, \text{atan}\left(\frac12 \sqrt{A_1^2+A_2^2} \right), \quad
\Omega = \atantwo(A_2,A_1), \notag \\
\omega &= \theta_p-\Omega, \quad a = \mu^{1/3}/A_6^{2/3},\label{eq:converse}\\
T(t) &= A_3(t)- \theta_p. \notag
\end{align}

Here $T(t)$ is the true anomaly at time $t$, so that
$\theta\tCRTN(t)$ in (\ref{eq:theta-crtn}) represents the remapped
angular direction of the orbital object with respect to the CRTN
preferred reference direction.  In addition $\theta_p\tCRTN$ in
(\ref{eq:thetap-crtn}) denotes the direction of perigee with respect to
the CRTN preferred reference direction, and $\phi_p\tCRTN$ in
(\ref{eq:phip-crtn}) denotes the mean anomaly version of this angle.
Finally $\phi\tCRTN(t)$ in (\ref{eq:phi-crtn}) is a mean anomaly
version of the information in $\theta\tCRTN(t)$.  Note that the 
initial central angles vanish
$\theta^{(c)\CRTN}(0) = \phi^{(c)\CRTN}(0) = 0$ but the deviated
initial angles $\theta\tCRTN(0) = F_\MtoT(\phi\tCRTN(0),e)$ are
small and nonzero.

The AST coordinates differ from equinoctial coordinates in three ways.
The most important difference is the choice of reference basis (the
CRTN basis for AST coordinates instead of the standard
ECI basis for equinoctial coordinates).  In particular, the CRTN basis
is \emph{adapted} to the distribution of the uncertain state whereas the
standard ECI basis is \emph{fixed}.  The other two differences are
that the angular position of the orbiting object, $A_3(t)$, is
represented on the mean anomaly scale rather than the true anomaly
scale (and is treated as a number rather than an angle), and that the
size of the ellipse, $A_6$, is defined by the mean motion $n$ instead
of the major semi-axis $a$.  These last two choices are made to
linearize the propagation equation (\ref{eq:mean-to-true}).

A detailed examination of AST coordinates will be given in the next
section.  For the moment attention is limited to some differences from
the standard equinoctial coordinates.

\begin{itemize}
\item[(a)] \textbf{Retrograde orbits.}  Standard equinoctial coordinates break
  down for a nearly retrograde equatorial orbit (for which the
  inclination approaches $180^{o}$).  For AST coordinates the problem does
  not arise since the inclination of the central state always equals $0^o$
  and the inclinations for the deviated states are always close to
  $0^o$.

\item[(b)] \textbf{Linear propagation and winding number.} The system
  equation (\ref{eq:phi-crtn}) is a linear function of
  $\phi\tCRTN(0)$ and $n$ for fixed time $t$.  Thus if the initial
  values of $\phi\tCRTN(0)$ and $n$ are Gaussian, the propagated
  value of $\phi\tCRTN(t)$ remains Gaussian for all future times
  $t$.

  Further, the use of this representation makes it straightforward to
  keep track of the \emph{winding number}, that is, how many times the
  orbiting object has gone around its orbit.  More specifically,
  without any knowledge of the history of the orbiting object, the
  initial angle $\phi\tCRTN(0)$ only makes sense as an angle; that
  is $\phi\tCRTN(0)$ and $\phi\tCRTN(0)+2\pi k$ represent the same
  angle for any integer $k$.  The initial angle can be turned into a
  number by restricting it to the interval $[-\pi,\pi)$.  Further
  since deviations from the central angle $\phi^{(c)}(0)=0$ are assumed
  small, the values of the deviated angles $\phi\tCRTN(0)$ are
  always close to 0.  Once $\phi\tCRTN(0)$ has been turned
  into a number, $\phi\tCRTN(t)$ in (\ref{eq:phi-crtn}) also makes sense as a number, and
  the integer part of $2\pi(\phi\tCRTN(t)-\phi\tCRTN(0))$ records
  the whole number of orbits which have occurred by time $t$.

\item[(c)] \textbf{Effects of rotation.}  Consider first a situation
  where the CRTN basis equals the standard ECI basis.  Hence the
  central orbital plane is equatorial and the initial central state
  points towards the standard reference direction,
  $\bm x^{(c)}(0) \propto \bm u^\std$.  Then the equinoctial and AST
  coordinates are identical except for small differences in elements 3
  and 6.  Next rotate the central and deviated orbital planes by
  $90^o$ to get polar orbits, but keeping the equinoctial reference
  frame unchanged.  The AST coordinates are unchanged.  However, since
  the reference plane for equinoctial coordinates remains equatorial,
  some of the equinoctial coordinates undergo major changes.
\begin{itemize}
\item[(i)] The third AST coordinate $A_3$ measures the difference in
  starting angular position along the orbital path between the
  deviated and central state.  However, for the polar orbit, the third
  equinoctial coordinate $E_3(0)$ also depends heavily on the RAAN
  angle $\Omega$.  As a result it is messy to interpret the angle
  $E_3(0)$ as a number near 0 and to develop the winding number
  interpretation in (b).

\item[(ii)] Suppose that before rotation the
  first two equinoctial coordinates, describing the deviation from
  vertical of the normal vector to the orbital plane, are
  approximately isotropic normal, $N_2(0, \sigma^2 I)$.  After
  rotation to a polar central orbit, the distribution of these
  coordinates is still approximately isotropic, but the variance
  becomes inflated (by a factor of 2 for a $90^o$ rotation).  The
  reason is $E_1(0)$ and $E_2(0)$  are defined by a stereographic
  projection of the sphere, and this projection is known to preserve
  the shape of a covariance matrix, but not its size~\cite{b21}.
\end{itemize}

\end{itemize}

\section {A first order representation for initial AST coordinates}
\label{sec:first} 
Let
\begin{equation}
R^{(c)} = [\bm u\tCRTN\ \bm v\tCRTN\ \bm w\tCRTN]
\end{equation}
denote the $3 \times 3$ rotation matrix constructed from the central RTN basis
using (\ref{eq:jtk-rtn-defn}).  Then \emph{standardize} all the deviated
states and the central state by defining
\begin{equation}
\label{eq:transformed-coord1}
\bm y(t) = R^{(c)T} \bm x(t), \quad \bm y^{(c)T}(t) = R^{(c)T} \bm x^{(c)}(t),
\end{equation}
\begin{equation}
\label{eq:transformed-coord111}
\bm{\dot{y}}(t) = R^{(c)T} \bm {\dot{x}}(t), \quad \bm{\dot{y}}^{(c)}(t) = R^{(c)T} \bm{\dot{x}}^{(c)}(t).
\end{equation}
After standardization, the
central state at time $t=0$ has coordinates
\begin{equation}
\label{eq:yc}
\bm {y}^{(c)}(0)  = 
\begin{bmatrix} A \\ 0 \\ 0 \end{bmatrix},
\quad
\bm{\dot{y}}^{(c)}(0)  = 
\quad
\begin{bmatrix} B \\ C \\ 0 \end{bmatrix}
\end{equation}
for parameters $A>0, B \in \mathbb{R}$, and $C>0$.  A deviated state can
then be written in the form 
\begin{equation}
\label{eq:yd}
\bm y(0)  = 
\begin{bmatrix} A+\epsilon_1 \\ \epsilon_2 \\ \epsilon_3 \end{bmatrix},
\quad
\bm{\dot{y}}(0)  =
\quad
\begin{bmatrix} B+\delta_1 \\ C+\delta_2 \\ \delta_3 \end{bmatrix}
\end{equation}
where $\bm \epsilon$ and $\bm \delta$ are typically ``small''.  In
this section, a first order Taylor expansion is used to show how 
the difference in  AST coordinates between the deviated and the
central state  can approximated by linear expressions of $\bm \epsilon$
and $\bm \delta$,  
\begin{equation}
\label{eq:Jacob-matrix}
\bm A - \bm A^{(c)} = 
\bm J \begin{bmatrix} \epsilon_1\\ \epsilon_2\\ \epsilon_3\\ 
\delta_1\\ \delta_2 \\ \delta_3 \end{bmatrix}.
\end{equation}
where $J$ is the 6$\times$6
Jacobian matrix from Cartesian-ECI to AST coordinates.

The formula for $J$ is derived in Appendix A.  The quality of this linear
approximation is explored in the next section.

\section {Linearity analysis for initial AST coordinates}
\label{sec:linearity}

This section explores numerically the extent to which AST coordinates
at the initial time $t=0$ depend linearly on $\bm \epsilon$ and
$\bm \delta$.  In order to simplify the study to its mathematical
essentials, suppose the length and time units are scaled so that the
gravitational constant is $\mu=1$ and the central orbital period is
$p=2\pi$.  Then an initial central state can be specified by giving
the eccentricity $e$ and the initial true anomaly $T(0)$.  The
corresponding values of $A,B,C$ are given by
\begin{align*}
A &= \frac{h^2}{1+e \cos T(0)}, \text{  where  } h^2 = 1-e^2,\\
C &= \frac{h}{A}, \quad B = \frac{e}{AC}\sin T(0).
\end{align*}

The error variances are most conveniently specified in terms of
relative errors.  For this study set the position standard error
$\sigma$ to be a specified percentage $P_\sigma\%$ of the geometric
mean of the radius at perigee and apogee.  Similarly set the velocity
standard error $\tau$ to be a specified percentage $P_\tau\%$ of the
geometric mean of the speed at the perigee and the apogee.  In
standardized units, these geometric means for position and velocity
reduce to $h$ and $1$, respectively.  More details are given in the
Appendix B.

For each component of $\bm \epsilon$ and $\bm \delta$, 7 equally
spaced values were chosen from $-2\sigma$ to $+2\sigma$ and $-2\tau$ to
$+2\tau$, respectively.  Then for each AST coordinate and each
coordinate of $\bm \epsilon$ and $\bm \delta$ a plot is constructed.
The plot shows how each AST coordinate varies as the corresponding
coordinate of $\bm \epsilon$ or $\bm \delta$ takes its 7 possible
values (with the other components of $\bm \epsilon$ and $\bm \delta$
fixed at 0).  Also superimposed on the plot is a straight line with
slope given by the corresponding element of the Jacobian matrix $J$.
Thus a total of 36 plots are generated.  If the mapping from
$\bm \epsilon$ and $\bm \delta$ is exactly linear, then the 7 ``test 
values'' in each of the 36 plots should lie exactly on a straight line.

\textbf{Example 1.} To judge the quality of the linear approximation
for AST coordinates, a challenging set of parameters was chosen, with
a high eccentricity, $e=0.7$, and high relative standard error,
$P_\sigma = 2.5\%$, $P_\tau = 20\%$.  This eccentricity is at the high
end of what is observed in practice.  The error rates are far higher
than usually found in practice, but are kept small enough to ensure
that deviated eccentricity is always less than 1.

Various choices were tried for the initial
true anomaly; the choice $T(0) = 45^{o}$ is shown here, but the choice of
$T(0)$ has little effect.

\begin{figure}[ht!]
\begin{center}
\includegraphics[width=16.4cm, keepaspectratio]{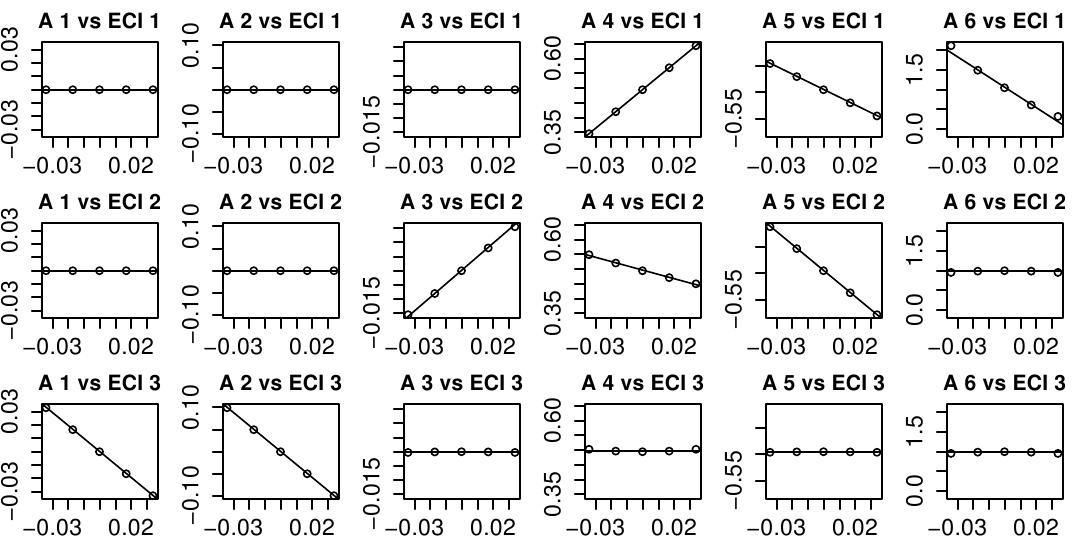}
\caption{\textbf{Example 1.} Linearity analysis at time $t=0$ showing
  plots of each AST coordinate against the first three ECI
  coordinates. See also Fig. \ref{fig600}.}
\label{fig60}
\end{center}
\end{figure}

\begin{figure}[ht!]
\begin{center}
\includegraphics[width=16.4cm, keepaspectratio]{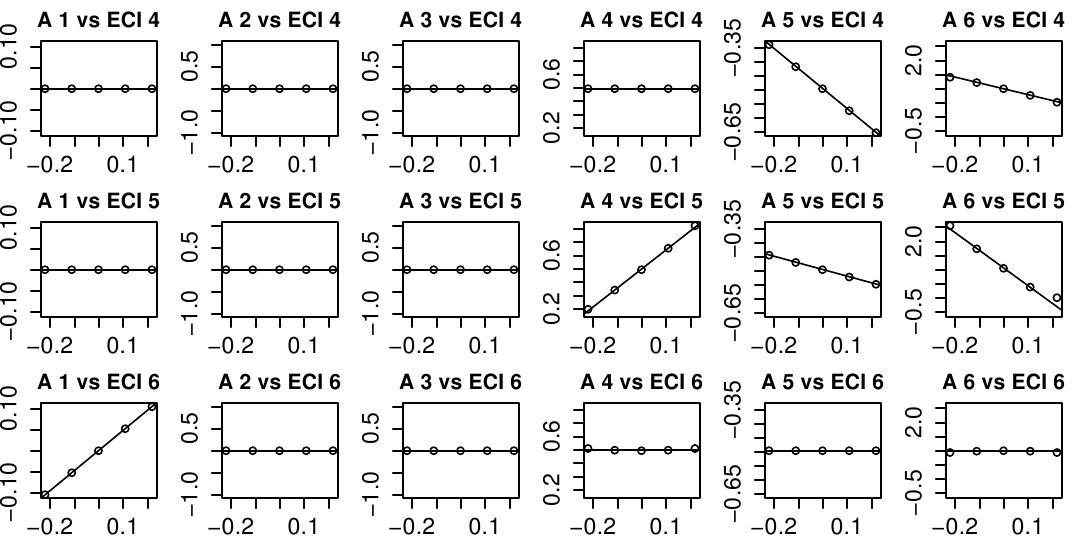}
\caption{\textbf{Example 1.} Linearity analysis at time $t=0$ showing
  plots of each AST coordinate against the last three ECI
  coordinates.}
\label{fig600}
\end{center}
\end{figure}

Figs. \ref{fig60}-\ref{fig600} show that the linearity approximation
is generally very good, even under these extreme conditions. Most of
the plots are visually very close to linearity.  The worst one is plot
(1,6) in Fig. \ref{fig60} with a squared correlation coefficient
of ``only'' $R^2 = 0.977$; even this plot is acceptably linear for most
purposes.  Of course the quality of the linear approximation improves
with lower standard errors.  It also improves with lower 
eccentricity.

Note that in Figs. \ref{fig60}-\ref{fig600}, within each row one ECI
coordinate varies over an interval with the other ECI coordinates held
fixed.  The rows in Fig. \ref{fig60} correspond to the three ECI
position coordinates.  The rows in Fig. \ref{fig600} correspond to the
three ECI velocity coordinates.

\section{Propagation  under different coordinate systems}
\label{sec:propagation}

Previous sections emphasized the initial behavior of deviated states
under different coordinate systems.  This section looks at the
propagated distributions after a given propagation time $t$ say, under
Cartesian-ECI, equinoctial and AST coordinates.

In Cartesian-ECI coordinates all 6 coordinates vary with time.
Further, their distribution is subject to the ``banana'' effect in
which uncertainty in the mean motion causes the distribution of the
propagated position to become spread out along the orbital path.
Equinoctial and AST coordinates are not susceptible to the banana
effect, in particular, only the third coordinate ($E_3(t)$ and $A_3(t)$
respectively) varies with time.

The following example illustrates some of the problems with
Cartesian-ECI and equinoctial coordinates.  Each 6-dimensional
propagated distribution is simulated and the resulting point cloud is
visualized using a pairs plot. Each pairs plot includes a histogram
for each variable and a scatter plot for each pair of variables.  The
point clouds are based $N=2000$ simulated initial states.  This value
of $N$ is more than sufficient to see the patterns of variability in
the propagated distributions.  Indeed the same patterns can be
identified using a much smaller value of $N$, e.g.  $N=500$.

\textbf{Example 2.} Consider a central orbit with eccentricity
$e^{(c)}= 0.7$ (an important parameter) and initial true anomaly
$T^{(c)}(0) = 45^o$ (a minor parameter).  Suppose the relative initial
standard deviations are $P_\sigma,P_\tau$, the  same as before.  For
equinoctial coordinates, the inclination is also an important
parameter.  If $i^{(c)}=0$, then equinoctial and AST coordinates are
very similar; here let $i^{(c)} = 158^o$ to illustrate the problems
that can arise for retrograde orbits.

In terms real world situations, if the period is 12 hours (equivalent to $a$ = 26610 km), these parameters correspond to
a highly eccentric orbit (HEO) with A = 9078 km, B = 2.6 km/sec and C = 8.1 km/sec. Further, $r_a$ = 45237 km, $r_p$ = 7983 km, $v_a$ = 1.6 km/sec and   
$v_p$ = 9.21 km/sec, where $r_a, r_p, v_a$ and $v_p$ indicate radius at the apogee, radius at the perigee, velocity at the apogee and 
velocity at the perigee respectively. 

\begin{figure}[ht]
\begin{center}
\includegraphics[width=16.4cm, keepaspectratio]{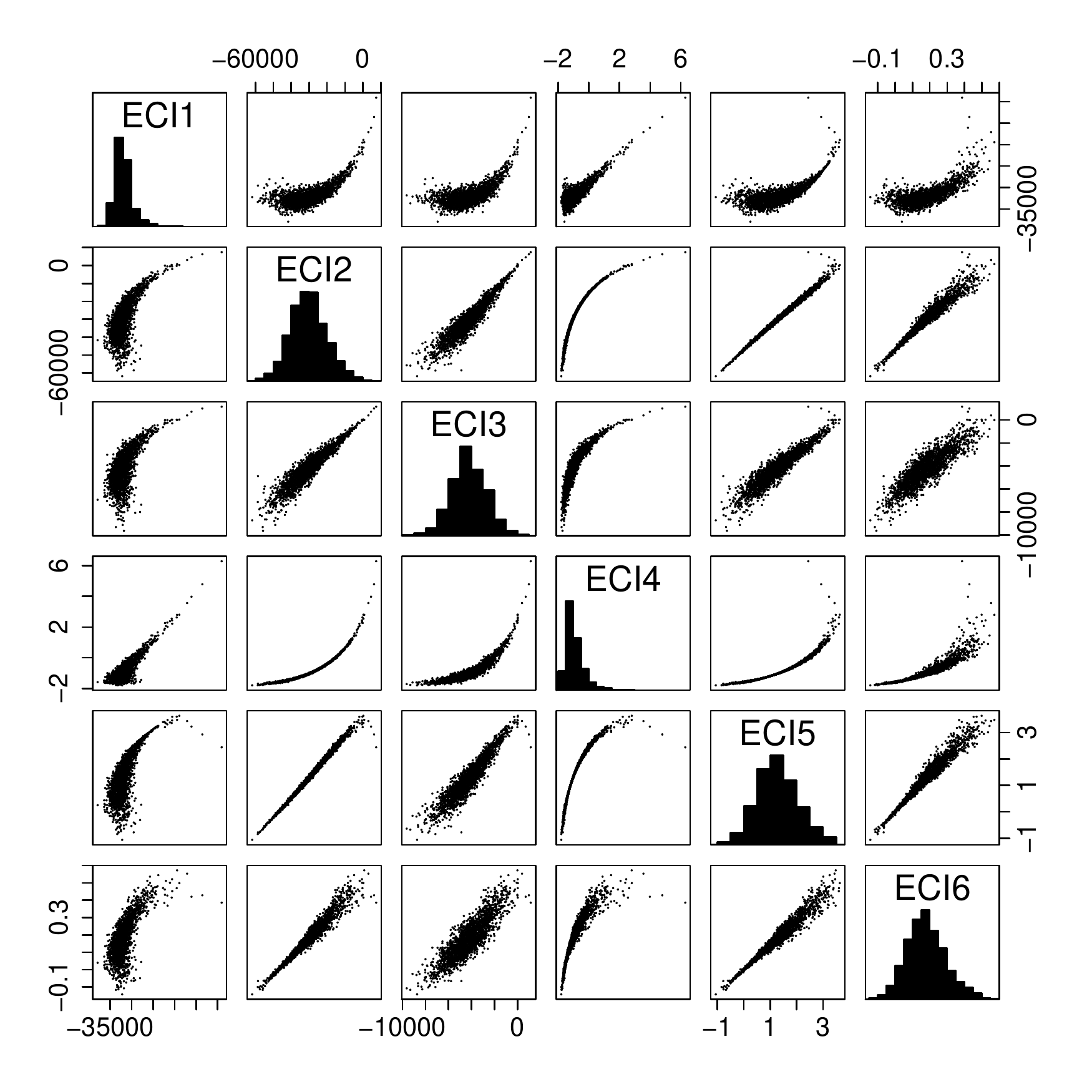}
\caption{\textbf{Example 2, Cartesian-ECI coordinates.} Propagated point cloud  in 
Cartesian-ECI coordinates. First three elements represent propagated position vectors (km) and last three elements indicate propagated 
velocity vectors (km/sec).}
\label{fig1}
\end{center}
\end{figure}

\begin{figure}[ht]
\begin{center}
\includegraphics[width=16.4cm, keepaspectratio]{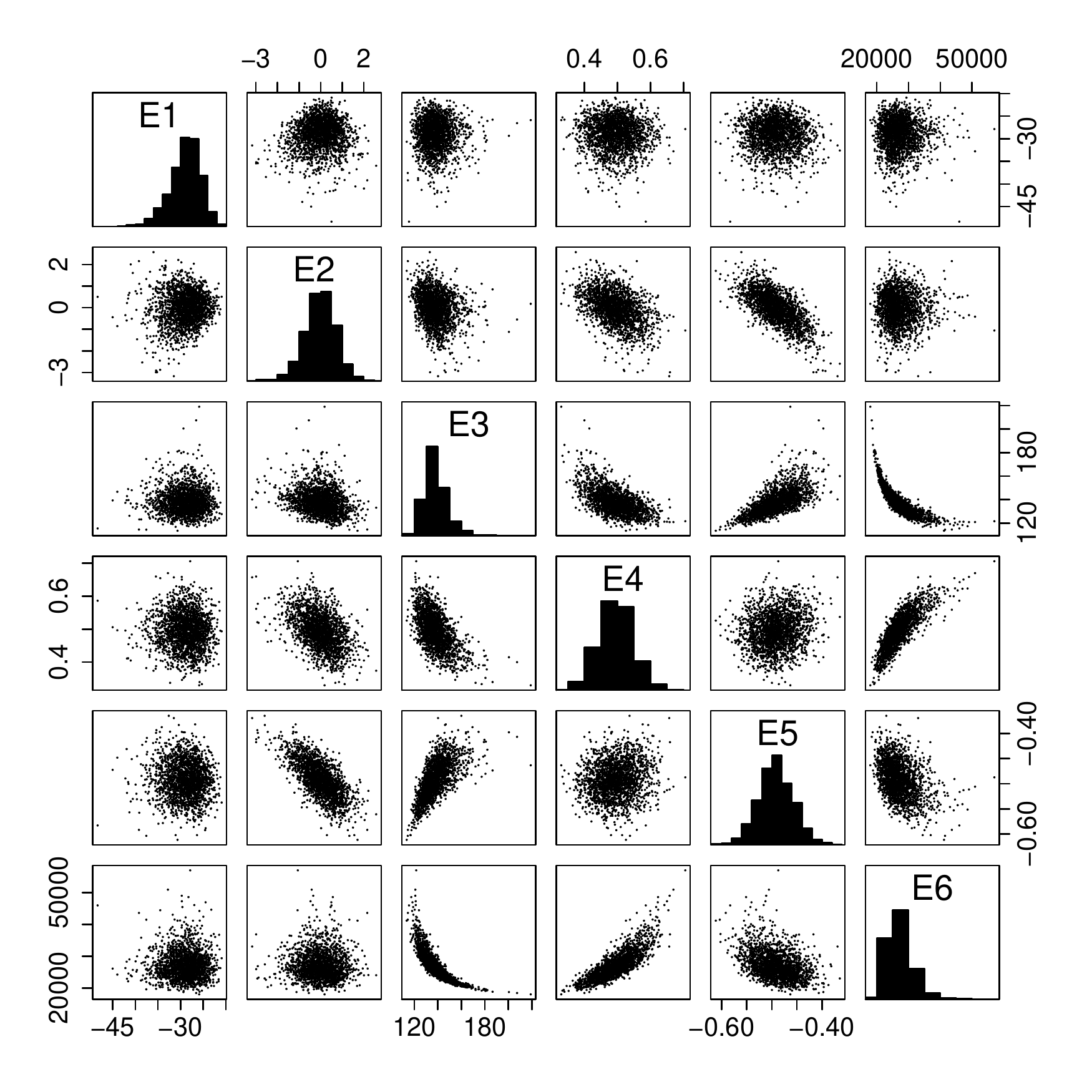}
\caption{\textbf{Example 2, equinoctial coordinates.} Propagated point cloud  in equinoctial coordinates.}
\label{fig3}
\end{center}
\end{figure}

\begin{figure}[ht]
\begin{center}
\includegraphics[width=16.4cm, keepaspectratio]{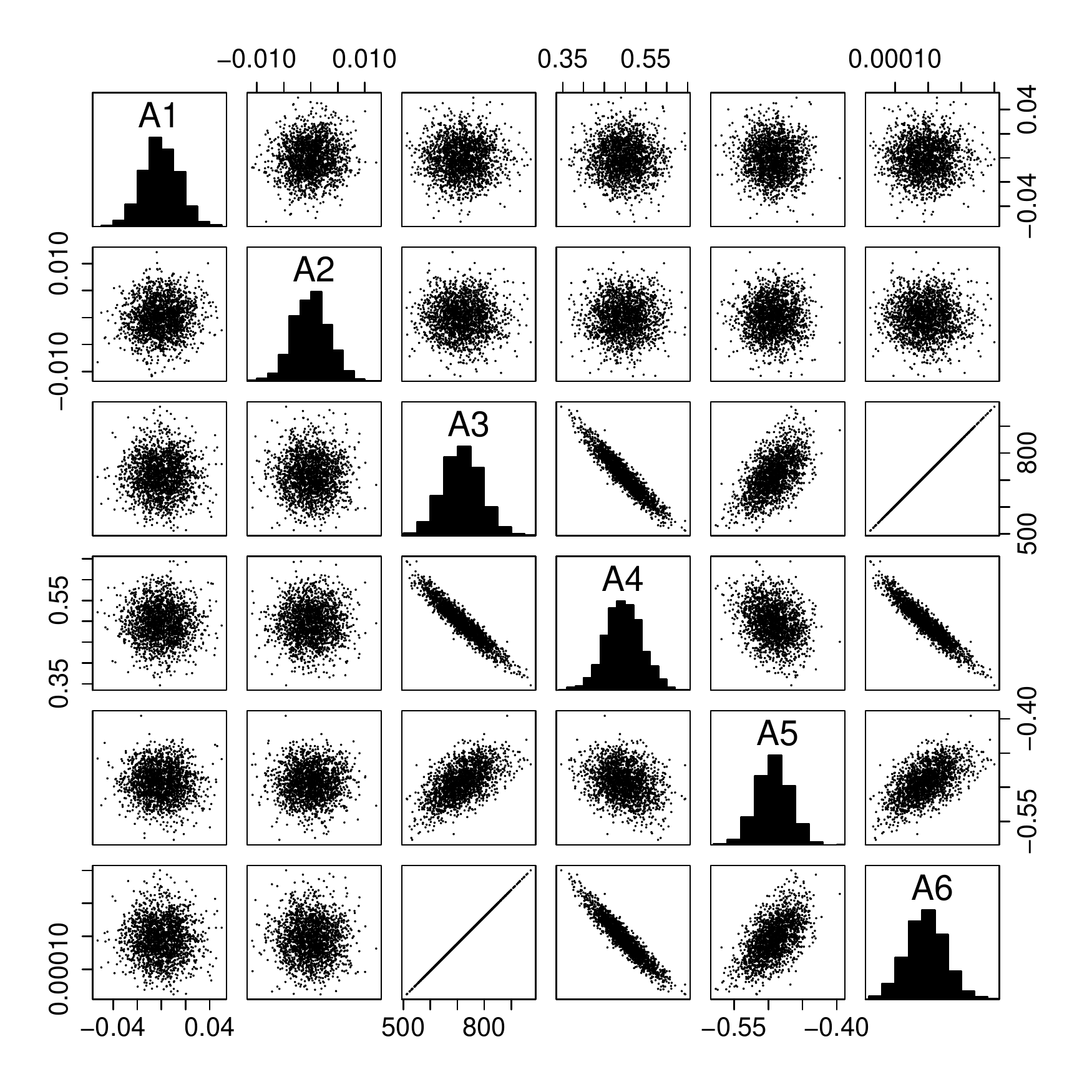}
\caption{\textbf{Example 2, AST coordinates.} Propagated point cloud in AST coordinates.  All the histograms and scatter plots are approximately normal.}
\label{fig51}
\end{center}
\end{figure}

The state of the object has been propagated for 0.5 central orbital
periods. Propagated point clouds have represented as 6-dimensional
pairs plot in Cartesian-ECI coordinates (Fig. \ref{fig1}), Equinoctial
coordinates with respect to the standard basis (Fig. \ref{fig3}) and
AST coordinates (Fig. \ref{fig51}).

From Figs. \ref{fig1}, \ref{fig3}, \ref{fig51}, the following
conclusions can be made.
\begin{itemize}
\item[(a)] In \textbf{Cartesian-ECI coordinates} (Fig. \ref{fig1}),
  there is extreme non-Gaussianity. E$.$g$.$ the scatter plot (2,4)
  shows severe curvature. Even with
  much lower standard deviation, there would still often be
  appreciable curvature in such plots.

\item[(b)] In \textbf{Equinoctial coordinates} (Fig. \ref{fig3}),
  $E_1$ there is also noticeable non-Gaussianity, e.g. the skewness in
  E1 and E2.  The non-Gaussianity in this example is due to the high
  inclination of the orbital plane and demonstrates the problems with
  equinoctial coordinates in this setting.

\item[(c)] In \textbf{AST coordinates} (Fig. \ref{fig51}), all the
  scatter plots are approximately normally distributed. Notice the
  perfect linear relation between elements 3 and 6 ($\phi(t)$ and
  $n$) in scatter plot (3,6), which is due to the fact that the
  uncertainty in $\phi(t)$ is dominated by the variability in $n$ for
  large $t$.  
\end{itemize}

It can be shown that non-Gaussianity can be even more severe for
Keplerian orbital elements~\cite{b17}.

\section{Statistical analysis for propagated distributions} 
\label{sec:stat_analysis}

An important criterion for a ``good'' coordinate system is that a point cloud at the
initial (and propagated) times should look approximately Gaussian,
given an initial  Gaussian distribution in Cartesian-ECI coordinates.
In the previous section Gaussianity was judged visually.  In this section
these judgments are backed up with formal statistical tests.
Two suitable test statistics were developed by Mardia~\cite{b8} to
assess skewness and kurtosis, respectively and have been implemented in R~\cite{b10}. 

For simplicity  attention is restricted to testing the full 6-dimensional
point cloud for Gaussianity under each of our coordinate systems.  The
results of each test can be summarized by a p-value. If Gaussianity
holds, the p-value will be uniformly distributed between 0 and
1. However, if normality fails, then the p-value will tend to be close
to 0. To carry out a formal statistical test, a small threshold
$\alpha$ is chosen (e.g. $\alpha=0.05$) and if the p-value is below
the threshold, then the hypothesis of Gaussianity is rejected. For
each pairs plot, two p-values have been computed, for skewness and
kurtosis, respectively.  The p-values for the \emph{Example 2} 
are summarized in Table \ref{my-label2}.  For both the Cartesian-ECI
and the equinoctial coordinate system p-values are very small. These
results reinforce our visual impression. 


\textbf{Caution}- The power of a statistical test depends on the
sample size (here the number of simulated points in the point
cloud). If the underlying distribution is even slightly non-normal,
then for a large enough sample size, the hypothesis of normality will
be eventually rejected. Here a sample size of $N = 2000$ has been
used, which is adequate to confirm the approximate normality in
Fig. \ref{fig51}, 
and to strongly reject normality in most of the plots in
Figs. \ref{fig1} and \ref{fig3}. In Table \ref{my-label2}, for the Cartesian-ECI and equinoctial coordinate systems 
very small p-values ($<$ 1e-20) are effectively 0 and indicate the distribution is extremely non-Gaussian.

\begin{table}[h!]
\centering
\caption{Normality test results. Here ${p_{\text{skewness}}}, {p_{\text{kurtosis}}}$ represent p-values for skewness and kurtosis 
respectively.} 
\label{my-label2}
\begin{tabular}{ccc}
\hline
Coordinate system  & ${p_{\text{skewness}}}$ & $ {p_{\text{kurtosis}}}$ \\
\hline

Cartesian-ECI (Fig. \ref{fig1})  & $<$ 1e-20 & $<$ 1e-20    \\ 
Equinoctial (Fig. \ref{fig3})  & $<$ 1e-20 & $<$ 1e-20    \\ 
AST coordinates (Fig. \ref{fig51}) & {0.27} & {0.4}   \\ \hline

\end{tabular}
\end{table}

\section{Filtering using the AST coordinate system} \label{filtering_Section}
\label{sec:filtering}

The main purpose behind the development of AST coordinates is to
facilitate the tracking of space objects. Assume that each observation
takes the form of an angles-only position measurement (a unit vector
$\bm z_\obs$), which can be represented using ``latitude''
$\psi_\obs  \in [-\pi/2,\pi/2]$ and ``longitude''
$\theta_\obs \in [-\pi,\pi)$, defined with respect to
the CRTN frame.  In general, a latitude
$\psi$ and longitude
$\theta$ determine unit vector $\bm z$ by
$$
z_1 = \cos \psi \cos \theta, \quad
z_2 = \cos \psi \sin \theta, \quad
z_3 = \sin \psi.
$$

\begin{figure}[ht!]
\begin{center}
\includegraphics[width=8.2cm, keepaspectratio]{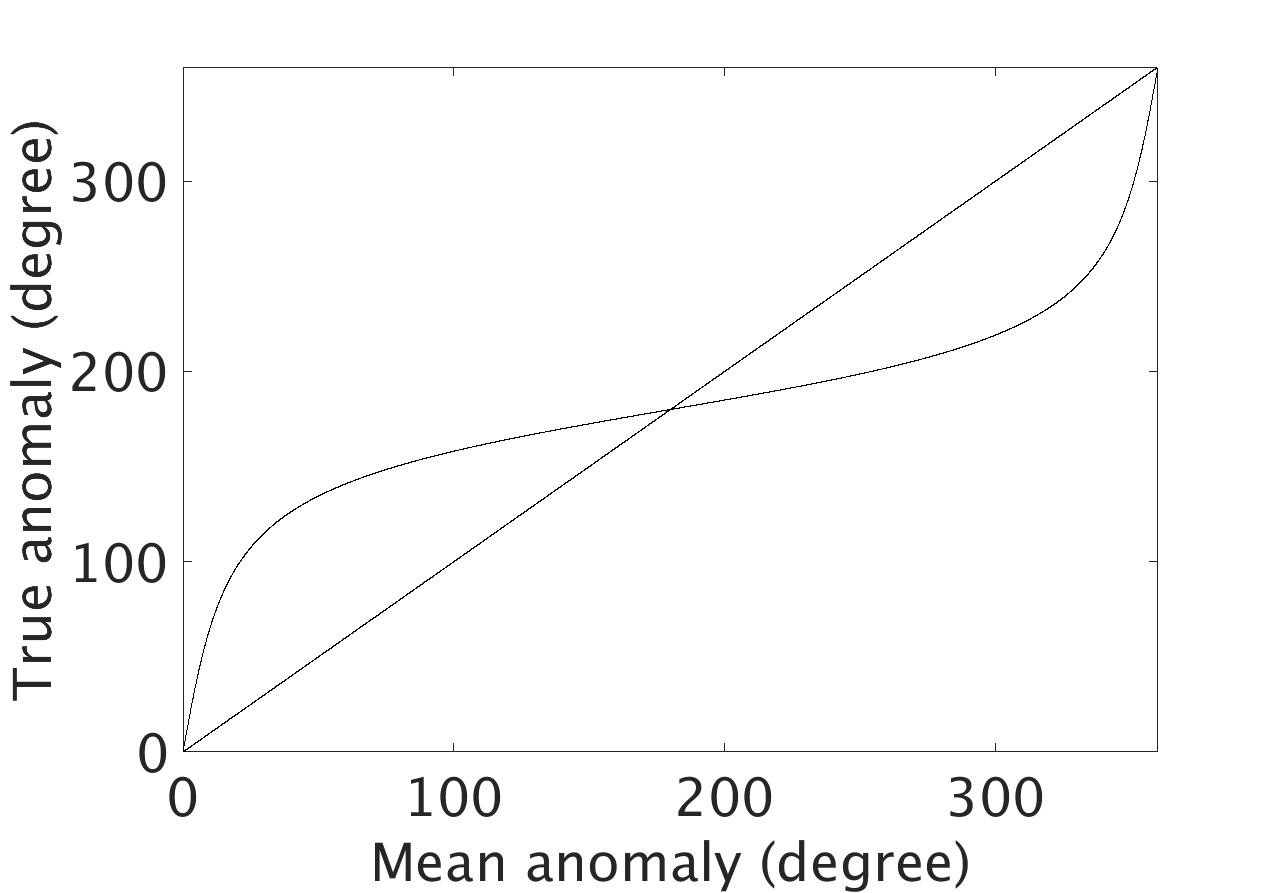}
\caption{\textbf{Mean anomaly to true anomaly.} The horizontal axis represents the mean anomaly and the vertical axis represents the true anomaly, both in degrees between $0^o$ and $360^o$.}
\label{fig_beg20}
\end{center}
\end{figure}

Longitude is measured on true anomaly scale (if there were no errors
the the observation would satisfy the identity
$\theta_\obs = \theta_p + T(t)$ would hold where $t$ is the
propagation time), whereas AST element 3 ($A_3(t) = \phi(t)$) is
computed on the mean anomaly scale. The transformation from mean
anomaly to true anomaly can be extremely non-linear under high
ellipticity.  Fig. \ref{fig_beg20} shows the relation when $e = 0.7$.

Next consider the update stage of the Kalman filter.  In particular,
it requires a 6-dimensional variance matrix for the propagated state
$\bm A(t)$, and a 2-dimensional measurement variance matrix for
$(\theta_{obs},\psi_{obs})$.  Of these the most interesting components
are the propagated variance of $A_3(t)$ and the measurement variance
of the longitude $\theta_{obs}$.  If the ellipticity is high, then a
nonlinear version of the Kalman filter is needed.  Common choices are
the unscented and extended Kalman filters (UKF and EKF)~\cite{b3, b11,
  b24, b19}.

However, if in addition the propagation time is large, then the
propagated variance of $A_3(t)$ can be much larger than the
measurement variance of $\theta_{obs}$.  In such a situation the UKF
and EKF can perform very poorly.  The reason is that they deal with
the nonlinearity by taking a first order Taylor expansion (discrete or
exact) centered at the propagated mean of $A_3(t)$, whereas it is much
better to center the first order Taylor expansion at or near the
measurement value $F_\TtoM(\theta_{obs},e)$, where $e$ is the
eccentricity for the mean propagated state.

These issues are discussed in more detail in a companion
paper~\cite{b23}, which makes 4 suggestions: (a and b) newly developed
observation-centered Kalman filters, (c and d) iterated Kalman filters
such as the IEKF~\cite{b14} and IUKF. This paper uses IUKF for update
steps. Note that one advantage of the OCEKF and OCUKF over the IEKF
and the IUKF is that they does not require iteration. However, an
advantage of the IEKF and the IUKF over the OCEKF and the OCUKF is
that they are more widely applicable.  In situations where measurement
error is not small, the posterior moments from the IEKF and IUKF can
be closer to the true posterior moments. Of course, a particle filter
can be used to solve the tracking problem, but it is not needed
here. Simple and very effective solutions are available using first
order approximations.

From equations (\ref{eq:ast})--(\ref{eq:phi-crtn}) and from Chapter 4
in \cite{b4} the spherical coordinates $(\theta(t),\psi(t))$ can be
written in terms of Keplerian elements as
\begin{align*}
\theta(t) &= \Omega + \omega +T(t),\\
\psi(t) &= \sin^{-1} \left\{\sin(L_1)\sin(L_2(t))\right\}, 
\end{align*}
where 
$L_1= 2\tan^{-1} i$ and $L_2(t) = \omega + T(t)$.
In turn, the Keplerian elements can be written in terms of the AST coordinates
using equation (\ref{eq:converse}).

\nil{
\begin{align*}
\theta_p &= \atantwo(f_2,f_1) = \atantwo(A_5,A_4),\quad
e = \sqrt{A_4^2+A_5^2},\\
 $\phi(t) = A_3(t)$ and
$\phi_p = F_\TtoM(\theta_p,e)$.  In addition, it follows from e.g. 
Chapter 4 in \cite{b4} that 
$$ 
\psi = \sin^{-1} \left(\sin(L_1)\sin(L_2)\right), \\
\end{align*}
where
\begin{align*}
L_1 = 2\tan^{-1} i, \quad
L_2 =  F_\MtoT \left(A_3(t),e \right) - \tan^{-1} \Omega \right).
\end{align*}
where $i = \sqrt{A_1^2 + A_2^2}/2$.
}

The AST-IUKF steps are briefly summarized below.\\
\fbox{\begin{minipage}{40em}
\begin{algorithm}[H]
\SetAlgoLined
\textbf{\underline{Given}}

a. The central state in Cartesian-ECI coordinates at time $t_0 = 0$\;
b. The covariance matrix associated with the central state in
Cartesian-ECI coordinate system\; c. Sequence of angles-only
measurements at times $t_1 <t_2 < \ldots$\;

\textbf{\underline{Computation}}

1) Find the CRTN frame at time $t_0$\; 2) Compute the initial mean
state and its covariance matrix in AST coordinates at time $t_0$, set
$\nu = 0$\; 3) Propagate the mean and variance of the AST state
distribution from time $t_\nu$ to time $t_{\nu+1}$\; 4) Transform the
angles-only measurements at time $t_{\nu+1}$ to CRTN coordinates\; 5)
Update the AST state mean and variance using the IUKF at $t_{\nu+1}$\;
6) Set $\nu \rightarrow \nu + 1$ and repeat stages 3), 4) and 5) for
each observation\;


\caption{AST-IUKF stages}
\label{algo1}
\end{algorithm}
\end{minipage}}

\vspace{5 mm}
\textbf{Note:} There is no need to update the CRTN frame at each iteration.  
 
 \subsection {Examples (filtering)}
This section analyses two examples. The first example (Example 3) illustrates the one-step update; the purpose is to 
show the limitations of the EKF and the UKF when the orbital eccentricity is high, the propagated variance is large and the observation variance is 
small.

The second example (Example 4) deals a more realistic tracking problem, using the IUKF algorithm along with 
the AST coordinate system to track a space object.

\textbf{{Example 3}, One-step update.} Assume an uncertain orbiting
object with central ellipticity $e^{(c)}=0.7$ and with initial
relative standard errors $P_\sigma =2.5\%,\ P_\tau=20\%$ in
Cartesian-ECI coordinates, the same as in Examples 1 and 2.  For
simplicity here assume the central inclination vanishes, $i^{(c)}=0^o$
and the central angle of perigee is $\theta^{(c)}_p =0^o$.  Recall
from (\ref{eq:mean-to-true}) that the propagated variance of $A_3(t)$
increases linearly with $t$.  Choose the propagation time $t_1$ large
enough that the standard deviation of $A_3(t)$ equals $\sigma^*=25^o$.
Also suppose that the propagated mean of $A_3(t)$ is $\mu^*=260^o$.
This value is chosen to highlight the nonlinearity of $F_\TtoM$.

Consider an angles-only observation with longitude $\theta_{obs}$ =
225.5$^{o}$ and latitude $\psi_{obs}$ = 0$^{o}$ in the CRTN frame of
reference, with measurement standard deviation 5.5e-04$^{o}$ (2
arc-seconds) for both. Note that the longitude of the observation,
after transformation to the mean anomaly scale, takes the value
$$
\phi_{obs} = F_\TtoM(225.5^{o}, 0.7) = 310^{o}, 
$$
which is located at the 2.5$\%$ upper tail of the propagated
distribution for $A_3(t_1)$ since $\phi_{obs} = \mu^* + 2
\sigma^* = 260^{o} + 2\times 25^{o} = 310^{o}$.  Such cases are
mildly unusual but not unlikely. 

The propagated distribution for $A_3(t_1)$ forms the prior in the
Bayesian update.  Since the measurement standard deviation is very
small (2 arc-seconds), the posterior mean for the $A_3(t_1)$ is
concentrated very close to 310$^{o}$. The nonlinearity of the function
$F_\MtoT$ between $260^o$ and $310^o$ leads to striking differences
between the various filters.  The results are summarized in Table
\ref{table:ex121}.  For the ``Exact'' entry in this table, 
the posterior mean and variance have been  computed using a particle filter
~\cite{b29, b30} with one million particles. Table
\ref{table:ex121} shows that that EKF and UKF give terrible approximations to the posterior 
distribution but the IUKF, IEKF, OCEKF and OCUKF all give excellent approximations.

The ``Exact'' columns gives the correct answer showing that the posterior
is highly concentrated about $310^o$.  The IUKF, IEKF, OCUKF and OCEKF
filters are all similar to one another and the exact answer.  However, the 
UKF and EKF are terrible filters.  Their 95\% probability intervals
are incompatible with the exact 95\% probability interval.

\begin{table}[t!]
\begin{center}
\caption{Posterior means and standard deviations for $A_3(t_1)$ in
  Example 3, computed using various filters.}
\label{table:ex121}
\begin{tabular}{cccccccccc}
\hline
Moment & UKF & IUKF & EKF & IEKF & OCUKF & OCEKF & ``Exact''\\ \hline
mean (A$_3$)     & 327.1$^{o}$   & 310$^{o}$     & 329.8$^{o}$       & 310$^{o}$ & 310$^{o}$ & 310$^{o}$ & 310$^{o}$ \\
s.d (A$_3$)     & 4.1e-04$^{o}$   & 3.2e-02$^{o}$     & 5e-04$^{o}$       & 3.1e-02$^{o}$ & 3.2e-02$^{o}$ & 3.3e-02$^{o}$ & 3.2e-02$^{o}$\\ \hline
\end{tabular} \end{center} 
\end{table}

\begin{figure}[ht!]
\begin{center}
\includegraphics[width=16.4cm, keepaspectratio]{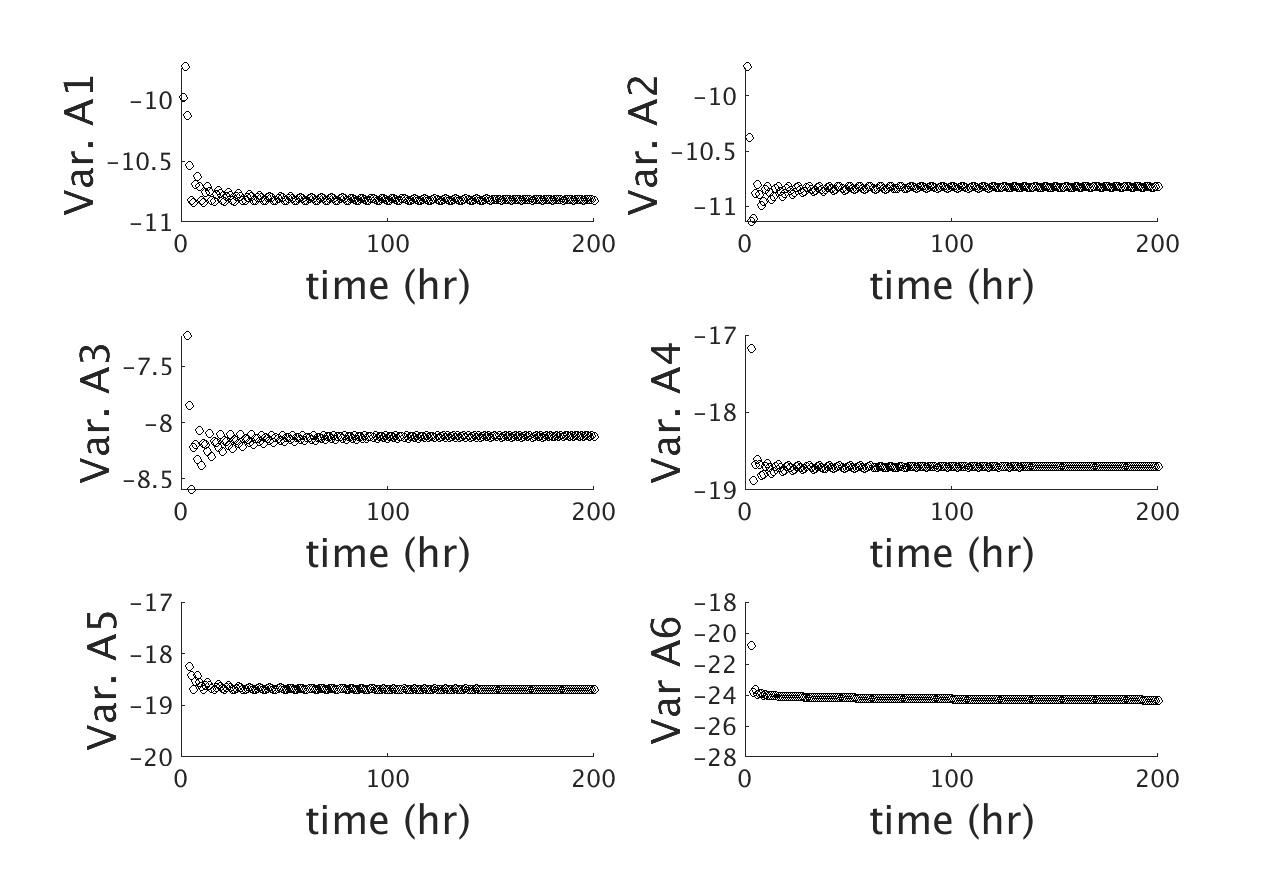}
\caption{\textbf{Example 4, Log scaled variance plots.} The log scaled updated AST variances vs$.$ time for A1-A6.}
\label{perfect112}
\end{center}
\end{figure}

\begin{figure}[ht!]
\begin{center}
\includegraphics[width=16.4cm, keepaspectratio]{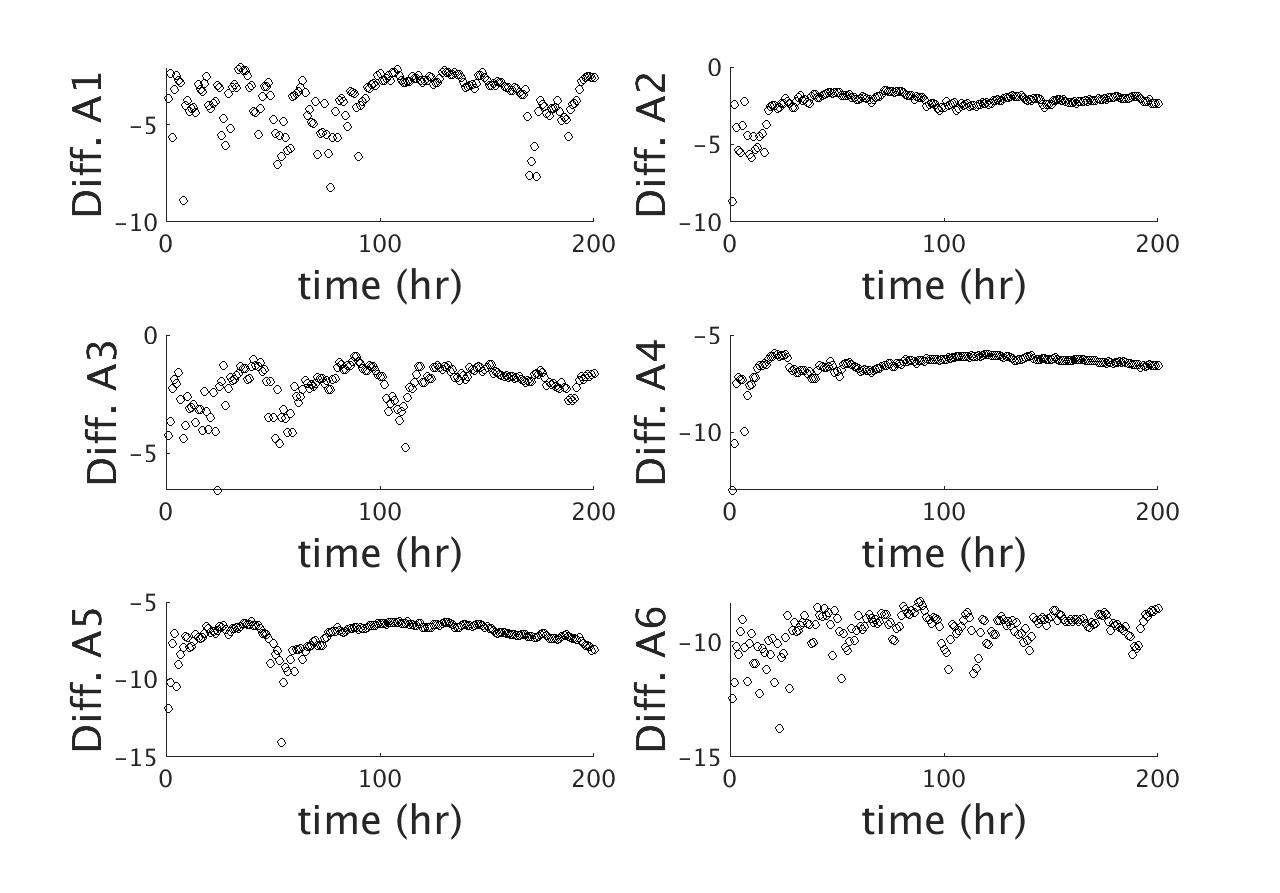}
\caption{\textbf{Example 4, Log scaled absolute difference plots.} The log scaled absolute differences between the true 
AST values and the updated AST means vs$.$ time for A1-A6.}
\label{perfect113}
\end{center}
\end{figure}

\textbf{{Example 4}, Tracking.} The purpose of this example is to
describe  effectiveness of the AST-IUKF algorithm for a sequence of
measurements. Consider the same  setup used in Example 2, and
consider a sequence of 200 hourly angles-only observations, with
standard deviations 0.1$^{o}$ in the  in-track and cross-track
directions. The results are summarized using Figs$.$ \ref{perfect112}
and \ref{perfect113}. In order to judge  the performance of the
AST-IUKF, two sets of plots have been generated and analyzed.  A brief
description is given below.  
\begin{itemize}
\nil{
\item[(a)] \textbf {Residual plots}. The $\nu^{th}$ \emph{residual} is
  the difference between the updated predicted and the observed
  angular positions at time $t_\nu$. Fig. \ref{perfect111} summarizes
  the residuals for longitude and latitude vs$.$ time. The residuals
  are mostly lie within $\pm$ 2 standard deviations (i.e. $\pm$
  0.2$^{o}$).   
  }
\item[(a)] \textbf {Log scaled variance plots}. Intuitively the AST
  posterior variances for $A_1(t_\nu)$ to $A_5(t_\nu)$ are expected to
  decrease at rate O($1/t_\nu$), and the posterior variance for
  $A_6(t_\nu)$ to decrease at rate O($1/t_\nu^2$). To visualize this
  behavior, Fig. \ref{perfect112} shows plots of
  $\log_{e}\{A_j(t_\nu)t_\nu\}, \ j=1, \ldots, 5$ and
$\log_{e}\{A_6(t_\nu)t_\nu^2\}$ vs$.$ $t_\nu$.  The log transform is used so
that a few initial outliers do not distort the plot. As expected, except for a few initial
values, each plot is approximately a horizontal straight line.   

\item[(b)] \textbf {Log scaled absolute difference plots}. Similarly,
  Fig. \ref{perfect113} shows plots of
  $\log_{e}\{D_j(t_\nu)t_\nu^{1/2}\}, \ j=1, \ldots, 5$ and
$\log_{e}\{A_6(t_\nu)t_\nu\}$ vs$.$ $t_\nu$, where $D_j(t_\nu)$ denotes the
absolute difference between the true AST value and the updated AST
mean at time $t_\nu$, for $j=1, \ldots, 6$.  
As expected, up to  sampling error all the plots  are
approximately horizontal straight lines.  
\end{itemize}

\section{Conclusion}
\label{sec:conclusion}

To summarize, this paper has investigated two types of nonlinearity.
The first type is the nonlinearity in the propagation equations in Cartesian-ECI coordinates.  
A first-order Taylor expansion was
used to represent the AST deviations at time $t = 0$ as approximately linear functions
of the ECI deviations. This expansion helps to explain why AST
coordinates are generally typically approximately Gaussian whenever
the initial conditions are. These conclusions are reinforced by the linearity plots in Section \ref{sec:linearity}, even under
extreme initial uncertainties.

The second type of nonlinearity arises from the difference between the true anomaly and mean anomaly
in the tracking problem. If the orbital eccentricity is high 
then standard tracking algorithms such as EKF or UKF sometimes fail to approximate the posterior mean and variance 
accurately. However, the iterated Kalman filters such as the IUKF, IEKF and the newly developed OCEKF and OCUKF work well in this situation.

\section*{Appendix}

\subsection* {A. The Jacobian for the mapping from Cartesian-ECI to
  AST coordinates.}  \label{app_ref1} The purpos of this appendix is
to show that the Jacobian $\bm J$ in (\ref{eq:Jacob-matrix}) between
Cartesian-ECI coordinates and AST coordinates at time $t=0$ takes the
form
\begin{equation*}
  \mathbf{J}=
  \begin{blockarray}{*{6}{c} l}
    \begin{block}{*{6}{>{$\footnotesize}c<{$}} l}
      $\epsilon_1$ & $\epsilon_2$ & $\epsilon_3$ &
      $\delta_1$ & $\delta_2$
      & $\delta_3$ &\\
\\
    \end{block}
    \begin{block}{[*{6}{c}]>{$\footnotesize}l<{$}}
      0 & 0 & -B/AC & 0 & 0 & 1/C &   \blank{0.5cm} \emph{A1}\\
      0 & 0 & -1/A & 0 & 0 & 0 &  \blank{0.5cm} \emph{A2} \\
      0 & D/A & 0 & 0 & 0 & 0 &  \blank{0.5cm} \emph{A3}\\
      C^2/\mu - 1/A & -BC/\mu & 0 & 
       0 & 2AC/\mu  & 0	& \blank{0.5cm} \emph{A4}\\
      -BC/\mu & B^2/\mu - 1/A & 0 & -AC/\mu & 
        -AB/\mu  & 0 &  \blank{0.5cm} \emph{A5}\\
      P_1C + P_2Q_{1} &  -P_1 B + P_2Q_{2} & 0 & P_1A + 2 P_2 A^2BC^2 & 
      P_2Q_{3} & 0  & \blank{0.5cm} \emph{A6}\\
    \end{block}
  \end{blockarray}
\end{equation*}

where
\begin{align*}
D &= \frac{\left(1-{e^{(c)}}^2\right)^{3/2}}
{\left(1 + {e^{(c)}}\cos T^{(c)}\right)^2}\\
P_1& = - \frac{3}{2}\frac{n^{(c)}}{{h^{(c)}}^2} 2AC \\
P_2 &= - \frac{3}{2}\frac{n^{(c)}}{{h^{(c)}}^2} \frac{a^{(c)}}{\mu} \\
Q_{1} &= \{(2C^2 - \frac{2\mu}{A})(AC^2 - \mu) + 2AB^2C^2\} \\
Q_{2} &= \{-2BC(AC^2 - \mu) - 2AB^3C + BC\mu\} \\
Q_{3} &= \{4AC(AC^2 - \mu) + 2A^2B^2C\}.
\end{align*}
Here is the derivation, with all expansions taken to first order in
$\epsilon$ and $\delta$.  The superscipt $^{(c)}$ denotes the value of
a parameter for the central state.  Recall that in $y$-coordinates
(\ref{eq:yc})--(\ref{eq:yd}), the CRTN basis is the same as the
  standard basis.

\textbf{Expansion for $A_1(0)$ and $A_2(0)$.}  The angular
momentum  vector can be expressed as
\begin{align*}
\bm h &= \bm x\times \bm{\dot{x}}\\
& =  \begin{bmatrix} 
\epsilon_2 \delta_3 - \epsilon_3(C+\delta_2) \\
\epsilon_3(B+\delta_1)-(A+\epsilon_1)\delta_3 \\
(A+\epsilon_1)(C+\delta_2) - \epsilon_2(B+\delta_1) \end{bmatrix}
\approx 
\begin{bmatrix}
-\epsilon_3 C\\ \epsilon_3 B - A \delta_3 \\ 
AC+A \delta_2 + C \epsilon_1 - \epsilon_2 B \end{bmatrix}
\end{align*}
with squared norm
$$
h^2 \approx  A^2C^2 +2AC (A \delta_2 + C \epsilon_1 - B \epsilon_2).
$$
That is,
$$
h^2 \approx h^{(c)2} + \Delta_{h^2},
$$
where $h^{(c)} = AC$ and
$$
\Delta_{h^2} = 2AC (A \delta_2 + C \epsilon_1 - B \epsilon_2).
$$
The first two components of $\bm w = \bm h/h$ simplify to
$$
w_1  \approx -\epsilon_3 C, \quad
w_2  \approx \epsilon_3 B - A \delta_3.
$$

In terms of Keplerian elements, 
$$
w_1 = \sin i \sin \Omega  \approx i\sin \Omega, \quad
w_2 = -\sin i \cos \Omega \approx -i\cos \Omega
$$
since the inclination angle $i$ is small.
Further, the first two AST coordinates are given by 
$$
2\tan(i/2)\sin(\Omega) \approx i \sin(\Omega) \approx w_1, \quad
2\tan(i/2)\cos(\Omega) \approx i \cos(\Omega) \approx -w_2,
$$
thus confirming the first two rows of $\/J$.

\textbf{Expansion for $A_4(0)$ and $A_5(0)$.} After a bit of calculation, the
expression for the eccentricity vector $\bm e$ simplifies to
$$
\bm e \approx \frac{1}{\mu} \begin{bmatrix} 
AC^2 + 2\delta_2 AC + \epsilon_1 C^2 - \epsilon_2 BC - \mu  -\mu \epsilon_1/A \\
-ABC - \delta_1 AC -\epsilon_1 BC - \delta_2 AB + \epsilon_2 B^2 
- \mu \epsilon_2/A \\
-\delta_3 AB -\epsilon_3 B^2 - \epsilon_3 C^2 -\mu \epsilon_3/A
\end{bmatrix},
$$
and since $\bm e = f_{1}\bm{u^{(c)}} + f_{2}\bm{v^{(c)}}$ from (\ref{eq:f}),  
\begin{align*}
\ f_1 &\approx \frac{1}{\mu} (AC^2 + 2\delta_2 AC + \epsilon_1 C^2 - \epsilon_2 BC - \mu  -\mu \epsilon_1/A),\\
\ f_2 &\approx \frac{1}{\mu} (-ABC - \delta_1 AC -\epsilon_1 BC - \delta_2 AB + \epsilon_2 B^2 
- \mu \epsilon_2/A).
\end{align*}
The first order error terms determine rows 4 and 5 of $\/J$.

\textbf{Expansion for $A_6(0)$.} The  squared norm of $\bm e$ becomes
\begin{align*}
e^2 \approx
& \frac{1}{\mu^2} \{ (AC^2-\mu)^2 + 2(2\delta_2 AC + \epsilon_1 C^2 
- \epsilon_2 BC -\mu \epsilon_1/A)(AC^2-\mu) \\
&+ (ABC)^2  2(\delta_1 AC + \epsilon_1 BC + \delta_2 AB - \epsilon_2 B^2 
+ \mu \epsilon_2/A)(ABC) \}\\
=& \{e^{(c)}\}^2 + \Delta_{e^2}
\end{align*}
where 
\begin{align*}
\{e^{(c)}\}^2 =&  \frac{1}{\mu^2} \{(AC^2-\mu)^2 + (ABC)^2 \},\\ 
{\Delta_{e^2}} =&\frac{1}{\mu^2} \{ 2(2\delta_2 AC + \epsilon_1 C^2 
- \epsilon_2 BC -\mu \epsilon_1/A)(AC^2-\mu)\\
 &+ 2(\delta_1 AC + \epsilon_1 BC + \delta_2 AB - \epsilon_2 B^2 
+ \mu \epsilon_2/A)(ABC)\}.
\end{align*}
 Then $a$ takes the form
\begin{align*}
a &= {\frac{h^2}{\mu}}{\frac{1}{1 -e^2 }}\\
\approx & {\frac{{h^{(c)}}^2}{\mu}}{\frac{1}{1 -{e^{(c)}}^2 }} (1 + 
\frac {{\Delta_h}^2}{{h^{(c)}}^2} +  \frac {{\Delta_e}^2}{1 -{e^{(c)}}^2})\\
&= {a^{(c)}} + \Delta_a,
\end{align*}
where
\begin{align*}
{a^{(c)}} & = \frac{{h^{(c)}}^2}{\mu}\frac{1}{1 -{e^{(c)}}^2 } \\
& = - \frac{A^2C^2}{A^2B^2C^2/\mu + A^2C^4/\mu - 2AC^2} \\
& = \frac{A\mu}{2\mu - AB^2- AC^2}.
\end{align*}
and
$$
\Delta_a = a^{(c)}(\frac {\Delta_{h}^2}{{h^{(c)}}^2} + 
\frac{\Delta_{e^2}}{1 -{e^{(c)}}^2}).
$$

The mean motion $n$ is the same as the sixth AST coordinate $A_6$ and
can be expressed as 
\begin{align*}
n &= (\mu/a^3)^{1/2} = \{\mu/(a^{c)} + \Delta_a)^3\}^{1/2}\\
&= (\mu/{a^{(c)}}^3)^{1/2}\{1 - \frac32 \Delta_a/a^{(c)}\}\\
&={n^{(c)}} + \Delta_n
\end{align*}
where $n^{(c)} = \{\mu/{a^{(c)}}^3\}^{1/2}$ and
\begin{align*}
\Delta_n &=  - \frac32 n^{(c)} \Delta_{a} /a^{(c)}\\
& = - \frac32 (n^{(c)}/a^{(c)}) 
\{\Delta_{h^2} / {h^{(c)}}^2 +  \Delta_{e^2} / (1 -{e^{(c)}}^2)\}\\
& = - \frac{3}{2}\frac{n^{(c)}}{{h^{(c)}}^2}(\Delta_{h^2} + 
\mu a^{(c)} \Delta_{e^2} ),
\end{align*}
since $1 / (1 -{e^{(c)}}^2) = \mu a^{(c)} / {h^{(c)}}^2$.  Simplifying
this expression yields the sixth row of $J$.

\textbf{Expansion for $A_3(0)$.}  For this section, write the first
order representation of the deviated basis at time $t=0$ in more
concise notation as
$$
\begin{bmatrix} \bm u & \bm v & \bm w \end{bmatrix} = 
\begin{bmatrix} 1 & -a & -b \\ a & 1 & -c \\ b & c & 1 \end{bmatrix},
$$
where $a=\epsilon_2/A, \ b=\epsilon_2/A, c = \delta_3/C-\epsilon_3 B/AC$.
The RANN direction $\Omega$ can be represented as a unit vector $\bm k$, say,
proportional to the vector cross product
$$
\bm w \times \bm w^{(c)} = \begin{bmatrix} -b \\ -c \\ 1 \end{bmatrix}
\times \begin{bmatrix} 0 \\ 0 \\ 1 \end{bmatrix} = 
\begin{bmatrix} -c \\ b \\ 0 \end{bmatrix}.
$$
After rescaling to a unit vector,
$\bm k = \begin{bmatrix} -c' \\ b' \\ 0 \end{bmatrix}$, where
$[b',c'] = [b,c]/\sqrt{b^2+c^2}$.

Rotate the initial deviated point $\bm y(0)$ to the equatorial plane
by the rotation $G$, say,  which keeps the node line fixed, and set
$\bm y^* = G \bm y$.  Then the angle $\theta(0)$ in
(\ref{eq:theta-crtn}) can be described as the angle in the equatorial
plane from $[1 \ 0 \ 0]^T$ to $\bm y^*$.

Write $G$ to first order in the form 
$$
G = I+Q = \begin{bmatrix} 1 & q_{12} & q_{13} \\ -q_{12} & 1 & q_{23} \\
-q_{13} & -q_{23} & 1 \end{bmatrix},
$$
where $Q$ is skew symmetric. The condition $G \bm k = \bm k$ implies that
$q_{12} = 0$ and $q_{13}c -q_{23}b = 0$.

The condition $R \bm w = [0 \ 0 \ 1]^T$ implies, after ignoring terms
higher than first order, that $q_{13}=b, \ q_{23} = c$.

Hence to first order 
$$
\bm y^* = R \bm y = \begin{bmatrix} 1 \\ a \\ 0 \end{bmatrix},
$$
and so  the angle from $\bm u^{(c)}$ is given by 
\begin{equation}
\label{eq:eps2}
\text{atan2}(a,1) = a = \epsilon_2/A. 
\end{equation}

The final step is to tranform from the true anomaly scale to the mean 
anomaly scale.  The value of $\theta(0)$ is related to the true
anomaly by $\theta(0) = \theta_p + T(0)$ and 
\begin{align*}
\phi(0) &= F_\TtoM(\theta_p,e)+F_\TtoM(T(0),e)\\
&= T_\TtoM(\theta(0)-T(0),e) -F_\TtoM(-T(0),e)\\
&\approx \theta(0) F'_\TtoM(-T(0),e) = \theta F'_\TtoM(T(0),e)\\
&\approx \theta(0) F'_\TtoM(T^{(c)}(0),e^{(c)}).
\end{align*}
The derivative is well-known (e.g. \cite{b4}),
\begin{equation}
\label{eq:T2Mder}
(d/dT)F_\TtoM(T,e) = \frac{(1-e^2)^{3/2}}{(1+e \cos T)^2}
= \frac{(1-e^2)^{3/2}}{(1+f_1)^2}.
\end{equation}
It does not matter to first order whether the deviated or central
value is used since $f_1$ and $e$ are close to $f_1^{(c)}$ and
$e^{(c)}$.  Combining (\ref{eq:eps2}) and (\ref{eq:T2Mder}) leads to
the third line of $\/J$.

\section*{B. Standardized units}
Section \ref{sec:linearity} refers to certain identities for orbital
elements when standardized units are used for length and time, so
that the gravitational constant is $\mu=1$ and the central orbital
period is $p=2\pi$.  From (\ref{eq:a})--(\ref{eq:p}) it follows that
$a=1$ and $h^2 = (1-e^2)$. Hence from \cite{b4} the standard formulas
for the radius at apogee and perigee, and the velocity at apogee and perigee,
simplify to 
\begin{align*}
r_a &= \frac{h^2}{\mu}\frac{1}{1+e\cos \pi} = \frac{h^2}{\mu}\frac{1}{1-e}
= 1+e, \\
r_p &= \frac{h^2}{\mu}\frac{1}{1+e\cos 0} = \frac{h^2}{\mu}\frac{1}{1+e}
= 1-e, \\
v_a &= \sqrt{\frac{(1-e)}{(1+e)}\frac{\mu}{a}} = \sqrt{\frac{(1-e)}{(1+e)}},\\
v_p &= \sqrt{\frac{(1+e)}{(1-e)}\frac{\mu}{a}} = \sqrt{\frac{(1+e)}{(1-e)}}.
\end{align*}
Therefore, geometric means become $\sqrt{r_a r_p} = \sqrt{1-e^2}$, 
$\sqrt{v_a v_p} = 1$.

\newpage
\nil{
\section*{C. AST variance and difference plots} \label{appendix-c}

\begin{figure}[ht!]
\begin{center}
\includegraphics[scale=0.42]{v2.jpg}
\caption{\textbf{Example 4, Variance plots.} Updated AST variances vs$.$ time for 
A1-A6.}
\label{perfect112-var}
\end{center}
\end{figure}

\begin{figure}[ht!]
\begin{center}
\includegraphics[scale=0.42]{D4.jpg}
\caption{\textbf{Example 4, absolute difference plots.} The absolute differences between the true AST values and 
the updated AST means vs$.$ time for A1-A6.}
\label{perfect113-abs}
\end{center}
\end{figure} 
}

\section*{Acknowledgment}
This material is based upon work supported by the Air Force Office of
Scientific Research, Air Force Materiel Command, USAF under Award
No. FA9550-19-1-7000.

\end{document}